\documentclass[aps,pre,twocolumn,groupedaddress,showpacs]{revtex4-1}
\usepackage{amsmath,amssymb,color}
\usepackage{epsfig}
\usepackage{xcolor}
\begin{document}
\def\red{\color{red}}
\def\E{\mathbb{E}}
\def\P{\mathbb{P}}
\def\R{\mathbb{R}}
\def\Z{\mathbb{Z}}
\def\O{{\cal O}}
\def\scr{\scriptstyle}
\def\text{\textstyle}
\def\floor{\rm floor}
\def\sw{\rm sw}
\def\al{\alpha}
\def\be{\beta}
\def\ga{\gamma}
\def\del{\delta}
\def\la{\lambda}
\def\sig{\sigma}
\def\om{\omega}
\def\Om{\Omega}
\def\phi{\varphi}
\def\na{\nabla}
\def\A{{\cal A}}
\def\atanh{\rm atanh}
\def\tiz1{\tilde z_1}
\def\ti{\tilde}
\def\p{\partial}
\def\sub{\subset}
\def\12{\frac{1}{2}}
\def\oraw{\overrightarrow}
\def\nea{\nearrow}
\def\sea{\searrow}
\def\beq{\begin{equation}}
\def\eeq{\end{equation}}
\def\beqn{\begin{equation*}}
\def\eeqn{\end{equation*}}
\def\beqa{\begin{eqnarray}}
\def\eeqa{\end{eqnarray}}
\def\beqan{\begin{eqnarray*}}
\def\eeqan{\end{eqnarray*}}
\title{Optimizing fog harvesting by biomimicry}
\author{J.C. Fernandez Toledano,  C. Fagniart, G. Conti, J. De Coninck}
\affiliation{Laboratoire de Physique des Surfaces et Interfaces\\
Universit\'{e} de Mons, 20 Place du Parc, 7000 Mons, Belgium}
\author{F. Dunlop} \author{Th. Huillet} 
\affiliation{Laboratoire de Physique Th\'{e}orique et Mod\'{e}lisation,
CNRS UMR-8089\\ CY Cergy Paris Universit\'{e}, 
95302 Cergy-Pontoise, France}
\begin{abstract}
Inspired by the stenocara beetle, we study an ideal
flat surface composed of a regular array of hydrophilic circular patches in
a hydrophobic matrix on an incline of tilt $\alpha$ with
respect to the horizontal. 
Based on an exact solution of the Laplace-Young equation at first order in the Bond number, the liquid storage capacity of
the surface is maximized as function of the patch radius, for suitable ranges of hydrophilic and hydrophobic contact angles, for tilt angles such as $45^{\circ}$ or $90^\circ$.
It is found that the optimal radius equally prevents dewetting from the top of the patches and overflow at the bottom. These theoretical considerations are validated by several experiments for the glass/octadecyltrichlorosilane (OTS) system involving different patch sizes and different inclinations. 
In a simple dynamical model, taking into account the flux of fog onto the surface or condensation on a suitably cooled surface, we find that the conditions for maximum harvest agree with the ones of maximum static storage.
The method could be developed for drop storage and drop transport  applications such as 
water-harvesting systems.

\end{abstract}
\maketitle
\section{Introduction}
Over 3,800 million years of evolution, Nature has given rise to optimal
structures that one can imitate for human purposes. Biomimicry studies
models of nature and imitates them or is inspired by them to solve a large
number of technological challenges. Some examples are adhesion inspired by gecko lizards \cite{Klittich17}, self assembly
DNA origami \cite{Praetorius17}, 
micropattern fabrication modelled after the \textit{salvinia molesta} \cite{Hunt11} or lotus-like self cleaning
materials \cite{Yu14}, among many others.

\begin{figure}[tbp]
\begin{center}
  \includegraphics[width=0.45\textwidth]{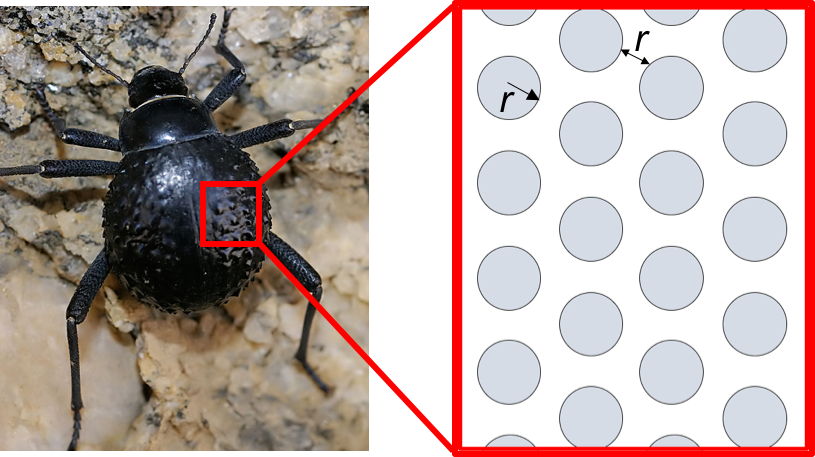} 
\end{center}
\caption{The stenocara beetle (photography by Hans Hillewaert under Creative
Commons), and a model of elytra as a triangular array of hydrophilic patches (grey disks) embedded in a hydrophobic matrix.}
\label{fig1}
\end{figure}

Biomimicry has been an interesting approach to increase
our understanding of dropwise condensation, which is fundamental to
water-harvesting systems \cite{Malik14}. In this sense, the stenocara beetle
shown in Fig. \ref{fig1}, surviving under extreme condition in the
Namibia desert, has been one of the archetypes of natural water harvest for
more than 15 years \cite{Parker01, Zhai06, Garrod07, Hong12}. These beetles
can get the water they need from dew collected from ocean fog, using their very own
body surfaces. Micro-sized bumps on the beetle's hardened forewings (\textit{%
elytra}) seem to help condense and direct water toward the beetle's awaiting
mandible. In early studies, the capacity of the stenocara to enhance water
condensation was associated to a combination of hydrophilic (water
attracting) bumps and hydrophobic (water repelling) areas. Indeed these
structures may increase fog- and dew-harvesting efficiency. For certain
species of Darkling beetle, the act of facing the foggy wind and
sticking its rear end up in the air (known as fog-basking behavior) is
thought to be just as important as body surface structure for successfully
harvesting water from the air (which has alternatively been suggested as
possibly being the beetle's alarm call behaviour \cite{Norgaard2010}).
However, more recent studies have reported that the entire elytra are
homogeneously covered with hydrophobic wax \cite{Norgaard2010, Guadarrama14}%
, questioning the role of surface chemistry in promoting condensation.
A new mechanism based on the role of the elytra surface
geometry has been proposed \cite{Park16}. In any case, inspired by the
initial model for the stenocara problem (hydrophilic patches surrounded by
a hydrophobic matrix), researchers started to design complex surfaces
combining regions with different wettabilities and special geometry for
efficient water guidance \cite{Zhai06, Garrod07, Hong12}. 

In this work the back of the beetle is modeled as a triangular array of 
hydrophilic circular patches of radius $r$ with hydrophobic spacing
also of order $r$, as sketched in Fig. \ref{fig1}, following the
estimation of the patch separation with respect to the patch radius from 
\cite{Garrod07}.
The choice of the triangular array is motivated by close-packing, maximizing the total patch area for given separation distance.
The size of the back of the beetle is fixed,
and so are the hydrophilic and hydrophobic contact angles. The beetle
inclines its back near the vertical, facing the foggy wind, and the water
collected is subject to gravity. The micro-droplets present in the air will reach the hydrophilic parts of the
surface (circular regions with a radius between 0.1 mm in initial works \cite
{Zhai06,Garrod07} to a few mm in more recent works \cite{Park16}) and some
of them will remain there to coalesce and form with time a bigger drop.
After some time, this sessile drop will cover some hydrophilic part of the
surface and will be in contact with the hydrophobic part. The first
question we ask: is there an $r$ which maximizes the water storage capacity, thus under statics conditions? The next question is about the dynamic behaviour: is harvesting from the fog wind or condensation related to this storage capacity?

These considerations have many potential applications first to enhance water collection which is interesting on its own but also for liquid transfer using micro-machines or for spotters in biotechnology. Recent promising applications for these heterogeneous wettability substrates deal also with enhancing the
energy-harvesting efficiency in the triboelectric nanogenerator or
hydroelectric generator systems \cite{triboelec};  site-selective chemical
reactions and catalysis;
 combining these wettability properties with
topological structures. Let us here point out that very interesting experimental results for condensation have been obtained recently proving the key importance not only of the patches but also of the patch sizes.
See \cite{Egab20, Song19} and \cite{Yang19}.

%%%%%%%%%%%%%%%%%%%%%%%%%%%%%%%%%%%%%%%%%%%%%%%%%%%%%%%%%%%%%%%%%%%%%%%%%%%%%%%%
Let us now present briefly our problem.
 Consider a drop with a circular basis on top of a patch, on an inclined surface. This drop will be characterized by two angles $\theta^{\min }$ at the top and  $\theta^{\max }$ at the bottom of the patch and by the Bond number $B=\rho g R^{2}/\gamma $, where $R$ is the radius of the spherical cap of same volume.  In the sequel, we shall also make use of the alternative Bond number $Bo= {\rho g V^{2/3}/\gamma} $, involving the volume of the drop.

 The maximum water storage of a single patch is reached when the bottom
contact angle $\theta^{\max }$ increases up to the advancing hydrophobic contact
angle $\theta_{2}$, leading to overflow, or when the top contact angle $\theta^{\min }$
decreases down to the receding hydrophilic contact angle $\theta_{1}$ , leading to dewetting. The angles satisfy 
(Fig. \ref{fig2}) 
\[
0<\theta_{1}\le \theta^{\min }<\theta^{\max }\le \theta_{2}<\pi 
\]
The angles $\theta^{\min }$ and $\theta^{\max }$ are of course functions of the size of the patch and drop volume. 
We will show in Section II that the drop with the largest volume is reached when precisely both conditions are satisfied : $\theta^{\min }=\theta_1$ and $\theta^{\max }=\theta_2$.
%This point is developed  in Sections 2. 

Moreover when we introduce the time in the problem to study for instance condensation or fog harvesting, we have to estimate different timescales: $t_1$, the time required to fill the patch with microdroplets, $t_2$, to fill the patch up to the maximum volume and $t_3$, for removal of the liquid from the surface typically within cascades. Let us point out that $t_3$ has to be small compared to $t_1$ and $t_2$ since we are dealing with mono-disperse array of patches.   This is considered in Section III. A series of experiments devoted to different patch sizes and different inclinations will be presented in Section IV. This will allow us to validate our theoretical predictions. A summary and discussion are given in Section V.

\section{Maximum static storage}
Let us denote $V(r)$ the maximum water storage of a single patch with radius 
$r$. The number of patches is proportional to $1/r^{2}$, so that $V(r)/r^{2}$ is,
up to a constant, the total maximum water storage of a collection of
hydrophilic patches arranged on a triangular lattice of surface $S$. Let us
compute this constant. The area of an hexagonal cell with side length $a$
(say in mm) is: $A=\left( 3\sqrt{3}a^{2}\right) /2,$ where $a=2r+\delta $
and $\delta $ is the distance between any two contiguous patches with radius $%
r$. In the sequel, we choose $\delta =r$. In a triangular array of patches,
there are $6\times 1/3+1=3$ patches per cell, so the density of patches (mm$%
^{-2}$) is:\emph{\ } 
\[
3/A=\frac{2}{\sqrt{3}\left( 2r+\delta \right) ^{2}}=\frac{2}{9\sqrt{3}r^{2}} 
\]
The number of patches on a pannel with surface $S$ is thus, neglecting boundary corrections: 
\begin{equation} \label{Numb}
N\left( r\right) =\left( 3S\right) /A=\frac{2S}{9\sqrt{3}r^{2}}, 
\end{equation}
so that the total volume storage on a pannel of surface $S$ is $N\left(
r\right) V\left( r\right) $. The factor in front of $V\left( r\right) /r^{2}$
is thus 
\begin{equation} \label{K}
K\left( S\right) =\frac{2S}{9\sqrt{3}}, 
\end{equation}
 and the total volume of
water stored on a surface $S$ is $K\left( S\right) V\left( r\right) /r^{2}.$

The question now is: does the function $V(r)/r^{2}$ reach a maximum at a
finite value of $r$?

For very small $r$ gravity is negligible, water forms a spherical cap of
contact angle reaching $\theta_2$ above each patch, of volume proportional
to $r^3$. Therefore the function $V(r)/r^2$ tends to zero as $r\to 0$. For
very large $r$ water accumulates near the bottom of the spherical cap,
forming a ribbon of cross-section $\mathcal{O}(1)$ and length $\mathcal{O}%
(r^{1/2})$. Therefore the function $V(r)/r^2$ tends to zero as $r\to\infty$.
As a result, it will reach at least one maximum between zero and infinity. One
could also optimize the shape of the hydrophilic patches, but it is likely
that the circular shape is optimal.

We have computed this maximum, in an approximation linear in the Bond
number, which was used in \cite{DDH17} to derive the formula 
\begin{equation}  \label{furmidge}
\gamma r{\frac{\pi}{2}}(\cos\theta^{\min}-\cos\theta^{\max})=mg\sin\alpha
\end{equation}
in the spirit of Furmidge \cite{F1962}. This formula, notably the
coefficient $\pi/2$, was shown in \cite{DDH17} to be within 1\% of almost
exact \textsl{Surface Evolver} simulations up to Bond numbers of the order
of 5, a range sufficient for the present study. 

We recall below the essential results of the first order approximation, in spherical coordinates from the start, as in \cite{DTDHS21}. 

%%%%%%%%%%%%%%%%%%%%%%%%%%%%%%%%%%%%%%%%%%%%%%%%%%%%%%%%%%%%%%%%%%%%%%%%%%%%%%%
\subsection{Exact solution at small Bond number in spherical coordinates}
%%%%%%%%%%%%%%%%%%%%%%%%%%%%%%%%%
\begin{figure}[tbp]
\begin{center}
  \includegraphics[width=0.45\textwidth]{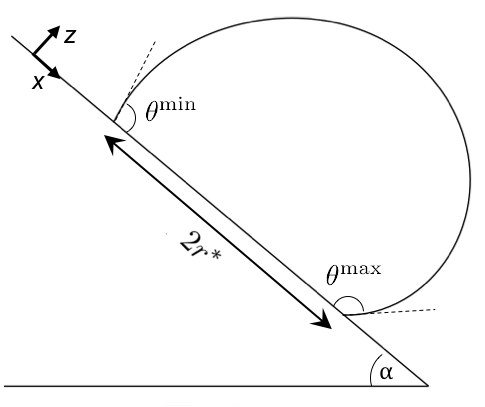} 
\end{center}
\caption{Drop profile maximizing $V(r)/r^{2}$ for $\theta_{2}=150^{\circ }$%
, $\alpha =40^{\circ }$}
\label{fig2}
\end{figure}
%%%%%%%%%%%%%%%%%%%%%%%%%%%%%%%%%
The pressure $p$ in the liquid, at any point on the interface, obeys both the
hydrostatic equation and the Laplace equation. Hence
\begin{multline}
  p_{\mathrm{atm}}-2\gamma H =p_{0}+\rho \,\mathbf{g}\cdot \mathbf{r}\\
  =p_{0}-\rho g\mathrm{r}\cos
\alpha \cos\theta+\rho g\mathrm{r}\sin \alpha \sin \theta\cos\varphi
\label{LY}
\end{multline}
where $p_0$ is the pressure at the origin, in the liquid,  and $H=(1/R_1+1/R_2)/2$ is the
mean curvature. The principal radii of curvature are negative when the
corresponding center of curvature is in the direction opposite the outer
normal (like the case of the spherical cap) and positive otherwise. We use
spherical polar coordinates with origin at the center of the spherical cap
at zero gravity, %for $B=0$, %see Fig. 4 in \cite {DDH17}, 
with $\theta\in [0,\theta_{0}]$ measured from the $z$-axis perpendicular
to the substrate and azimuth $\varphi \in [-\pi ,\pi ]$ measured from the $x$%
-axis downhill (see Fig. \ref{fig2}). So $(\mathrm{r},\theta,\varphi )$ in (%
\ref{LY}) are the coordinates of a running point on the drop surface.

We are interested in the deformation of the spherical cap due to gravity,
which is encapsulated in the last two terms of (\ref{LY}). Linear response suggests
to look for solutions to the Young-Laplace equation (\ref{LY}) having  the same form as functions of $\varphi$, namely
\begin{multline}
\mathrm{r}(\theta,\varphi )=R\Bigl( 1+Br_{01}(\theta)\cos\alpha
+Br_{11}(\theta)\sin \alpha \cos\varphi\cr
+O(B^{2})\Bigr)  \label{ansatz}
\end{multline}
where $r_{01}(\theta)$ and $r_{11}(\theta)$ are unknown functions. Then the
partial differential equation (\ref{LY}) reduces to ordinary differential
equations for $r_{01}(\theta)$ and $r_{11}(\theta)$. These turn out to be exactly
solvable:
\beq%\label{r01}
  r_{01}(\theta)={\cos\theta-\cos\theta_0\over 6}
  +\frac{\cos\theta}{3} \log{1+\cos\theta_0\over 1+\cos\theta}\label{r01}
  \eeq
  \beq%\label{r11}
  r_{11}(\theta) ={\sin\theta\over 3}\left[{\cos\theta\over 1+\cos\theta}
-{\cos\theta_0\over 1+\cos\theta_0}+\log{1+\cos\theta\over 1+\cos\theta_0}\right]\label{r11}
  \eeq
  Plugging (\ref{r01})(\ref{r11}) into (\ref{ansatz}) and dropping $O(B^2)$,
we compute exactly the contact angle $\theta(\varphi)$ along the contact line,
  \beq
\cos\theta={\cos\theta_{0}+B\,X(\theta_0,\varphi)\sin \theta_{0}\over
(\,1+B^2\,X^2(\theta_0,\varphi)\,)^{1/2}
}  \label{costheta}
\eeq
with
\beq
X(\theta_0,\varphi)=r_{01}'(\theta_0)\cos\alpha+r_{11}'(\theta_0)\sin\alpha\cos\varphi
\label{X}
\eeq 
\beq
r_{01}'(\theta_0)=-\sin\theta_0/6+\sin\theta_0\cos\theta_0/(3(1+\cos\theta_0))
\eeq
\begin{equation}\label{r'11}
r_{11}'(\theta_0)=\cos\theta_0/3-2/(3(1+\cos\theta_0))
\end{equation}

For comparison with simulation or experiment we shall make use of the Bond number $Bo$ obtained using the volume of the drop,
\beq
Bo={\rho
gV^{2/3}\over\gamma}=B\left[ \frac{\pi }{3}\left( 1-\cos \theta _{0}\right) ^{2}\left( 2+\cos
\theta _{0}\right) \right] ^{2/3}
\eeq
For example a hemispherical water drop with $R=3\,$mm has $B\simeq 1.25$ and
$Bo\simeq2.05$.
%%%%%%%%%%%%%%%%%%%%%%%%%%%%%%%%%%%%%%%%%%%%%%%%%%%%%%%%%%%%%%%%%%%%%%

Just before roll-off,
$\theta^{\max }=\theta_{2}$, or $\theta^{\min }=\theta_{1}$.
Let us start with the scenario $\theta^{\max }=\theta_{2}$, overflow at the front, which we label with a subscript 2. 
Taking the square of (\ref{costheta}) yields a second degree equation allowing to get
the unknowns $B_2$ and $R_2$ as functions of the single unknown $\theta_{0}$: 
\begin{equation} \label{bond}
B_{2}\left( \theta _{0}\right) =\frac{\sin \left( 2\theta _{0}\right) \pm
\sin \left( 2\theta _{2}\right) }{2X(\theta_0,0)\left( \cos ^{2}\theta _{2}-\sin
^{2}\theta _{0}\right) }.
\end{equation}

\begin{equation}\label{eq:bond}
R_2\Bigl( \theta_{0}\Bigr)=({%
\frac{\gamma B_2\left( \theta_{0}\right) }{\rho g}})^{1/2}
\end{equation}

The equation for $B_2$ can be expressed in a more simple way as
\begin{equation}  \label{tan2}
B_{2}\left( \theta _{0}\right) =\frac{\tan \left( \theta _{0}\pm \theta_2\right) }{X(\theta_0,0) }.
\end{equation}

The two solutions associated with the $\pm$ choices solve the square of (\ref{costheta}) equation, but we must examine whether they solve (\ref{costheta}), or perhaps (-\ref{costheta}) obtained from (\ref{costheta}) by adding a minus sign in front. Inserting (\ref{bond}) into (\ref{costheta}) gives
\beq
{\cos(\theta_2-\theta_0)\cos(\theta_2+\theta_0)\over|\cos(\theta_2-\theta_0)\cos(\theta_2+\theta_0)|}
={\cos(\theta_2\mp\theta_0)\over|\cos(\theta_2\mp\theta_0)|}
\eeq
Therefore (\ref{bond}) with choice + solves (\ref{costheta}) if and only if
$\cos(\theta_2+\theta_0)\ge0$, and (\ref{bond}) with choice - solves
(\ref{costheta}) if and only if $\cos(\theta_2-\theta_0)\ge0$. Note that
(\ref{costheta}) for each $\theta,\,\theta_0$, may have 0,\,1 or 2 solutions
for the unknown $B$.
In practice, we always have $\cos(\theta_2+\theta_0)<0$ and $\cos(\theta_2-\theta_0)>0$, which implies that '-' is the one valid choice.

The volume of the drop is given by the spherical cap formula 
\begin{equation}  \label{VoverR2}
V_{2}\left( r\right) =\pi r^{3}\frac{\left( 1-\cos\theta _{0}\right)
^{2}\left( 2+\cos \theta _{0}\right) }{3\sin ^{3}\theta _{0}}
\end{equation}
with $r=R_2\left( \theta_{0}\right) \sin \theta_{0}$, so that 
\begin{equation} \label{V:r2}
\frac{V_{2}\left( r\right) }{r^{2}}=\pi R_{2}\left( \theta _{0}\right) \frac{%
\left( 1-\cos \theta _{0}\right) ^{2}\left( 2+\cos \theta _{0}\right) }{%
3\sin ^{2}\theta _{0}}
\end{equation}
a function of $\theta_{0}$ only whose maximum is readily found.
The range of $\theta_0$ is restricted by $\theta^{\min }\ge\theta_{1}$,
specified by (\ref{costheta}) with $\varphi=\pi$.

Similarly, in the scenario $\theta^{\min }=\theta_{1}$, with $\varphi =\pi $, dewetting at the back, we get 
\begin{equation}
B_{1}\left( \theta _{0}\right) =\frac{\sin \left( 2\theta _{0}\right) \pm
\sin \left( 2\theta _{1}\right) }{2X(\theta_0,\pi)\left( \cos ^{2}\theta _{1}-\sin
^{2}\theta _{0}\right)}
\end{equation}

and $B_1$ can be expressed as
\begin{equation} \label{tan1}
B_{1}\left( \theta _{0}\right) =\frac{\tan \left( \theta _{0}\pm \theta_1\right) }{X(\theta_0,\pi) }.
\end{equation}
The choice + is allowed if $\cos(\theta_1+\theta_0)\ge0$, and the choice - is
allowed if $\cos(\theta_1-\theta_0)\ge0$.
In practice, we generally have $\cos(\theta_1+\theta_0)<0$ and $\cos(\theta_1-\theta_0)>0$, which forces '-' as the one valid choice.

This leads to $V_1(r)/r^2$ as a function of $\theta_{0}$ only whose maximum is also readily found.
The range of $\theta_0$ is now restricted by $\theta^{\max}\le\theta_{2}$,
specified by (\ref{costheta}) with $\varphi=0$.

Except for super-hydrophobic or super-hydrophilic materials, it turns out
that the maxima lie at the boundary of the ranges, where
$\theta^{\max }=\theta_{2}$ and $\theta^{\min }=\theta_{1}$ simultaneously.
Therefore for ordinary materials the optimal value of $\theta_{0}$ will be found while solving $B_{1}\left(
\theta_{0}\right) =B_{2}\left( \theta_{0}\right) ,$ when overflow and
dewetting occur simultaneously.

%%%%%%%%%%%%%%%%%%%%%%%%%%%%%%%%%%%%%%%%%%%%%%%%%%%%%%%%%%%%%%%%%%%%%%%%%%%%%%%

\subsection{Results : $\alpha =90\,$degrees}

%%%%%%%%%%%%%%%%%%%%%%%%%%%%%%%%%

\begin{figure}[tbp]
\begin{center}
  \includegraphics[width=0.45\textwidth]{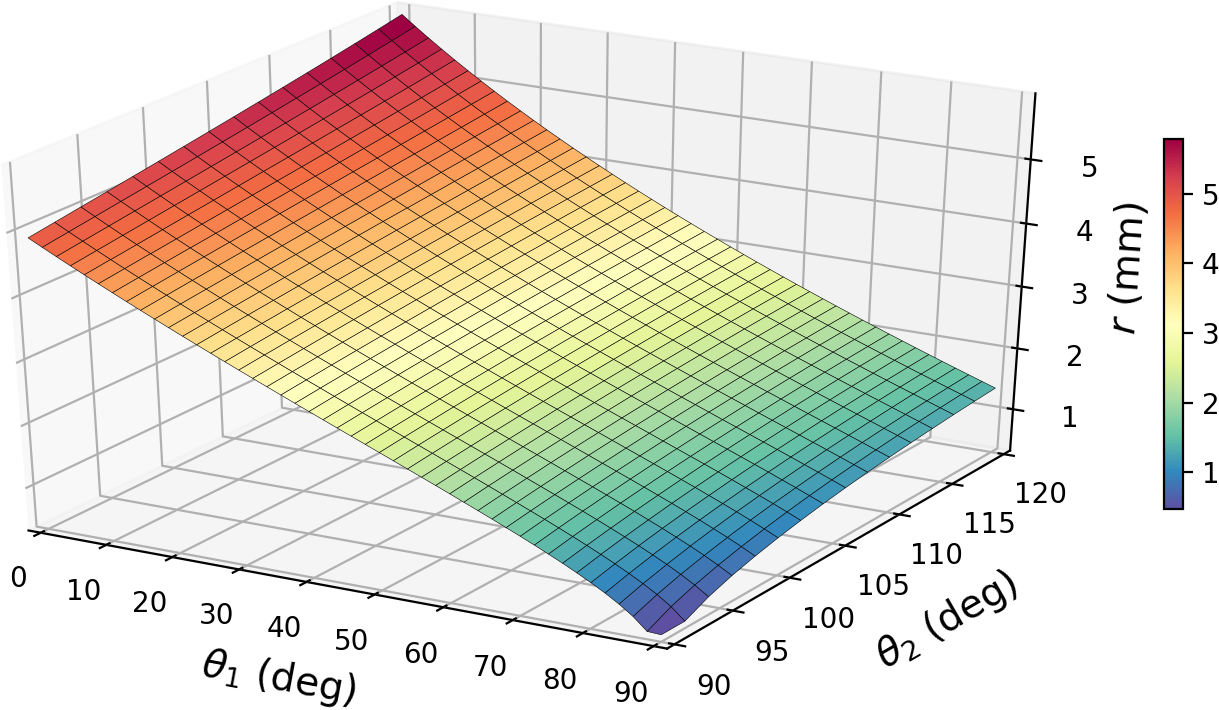} 
\end{center}
\caption{Optimal radius function of $\theta_1$ and $\theta_2$ when the optimum is for simultaneous dewetting and overflow. Slope $\alpha=90\,$degrees.}
\label{cond}
\end{figure}

%%%%%%%%%%%%%%%%%%%%%%%%%%%%%%%%%

%%%%%%%%%%%%%%%%%%%%%%%%%%%%%%%%%

\begin{figure}[tbp]
\begin{center}
  \includegraphics[width=0.45\textwidth]{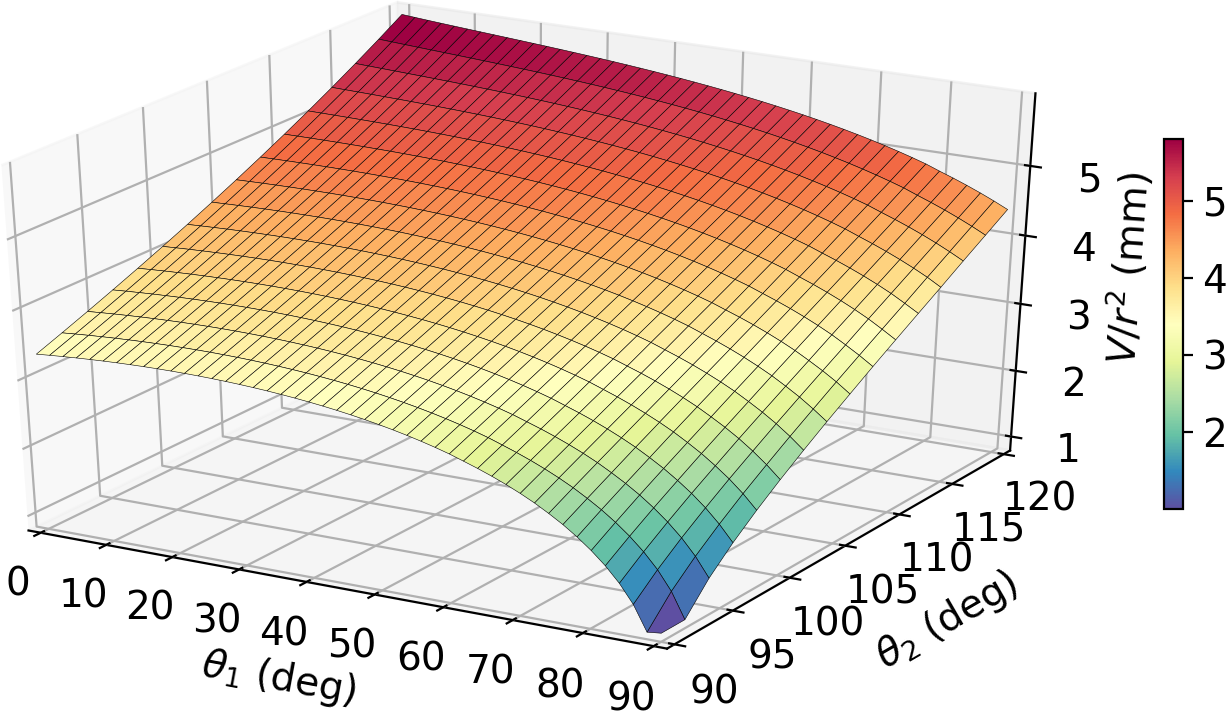} 
\end{center}
\caption{Optimal $V(r)/r^2$ function of $\theta_1$ and $\theta_2$ when the optimum is for simultaneous dewetting and overflow. Slope $\alpha=90\,$degrees.}
\label{Vr2}
\end{figure}

%%%%%%%%%%%%%%%%%%%%%%%%%%%%%%%%%

%%%%%%%%%%%%%%%%%%%%%%%%%%%%%%%%%
\begin{figure}[tbp]
\begin{center}
\resizebox{8cm}{!}{\includegraphics{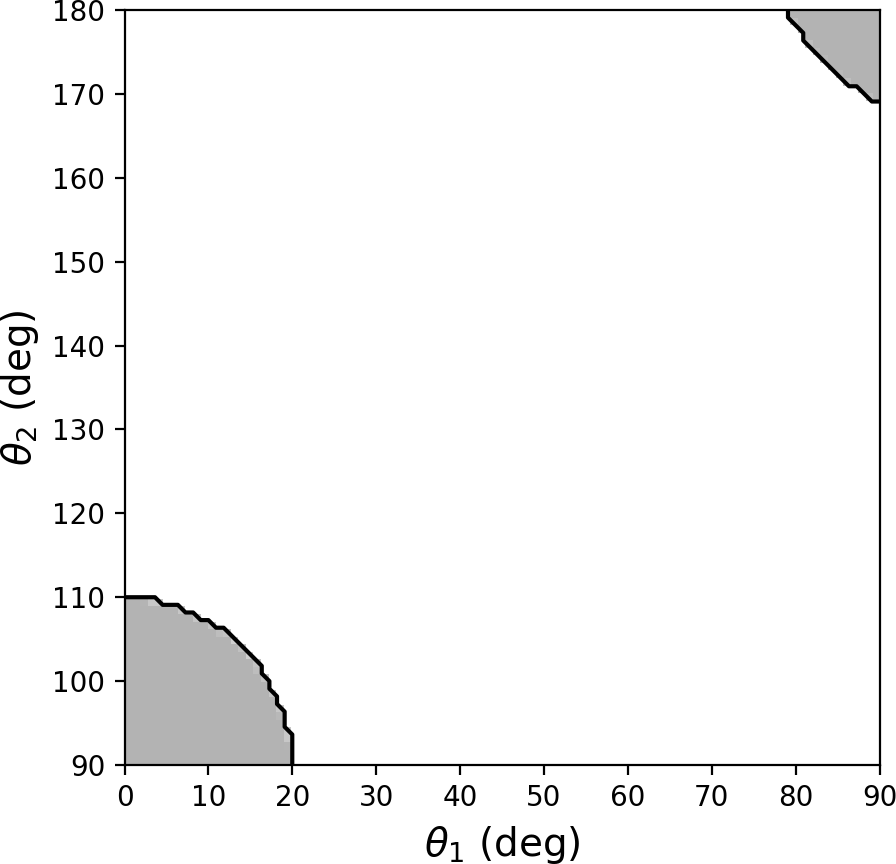}}
\end{center}
\caption{Line of equation (\ref{line}).
In the grey regions, the optimal radius of the patch is found on the overflow curve. In the white region, it is found when simultaneous dewetting and overflow apply. Slope $\alpha=90\,$degrees.}
\label{cont}
\end{figure}
%%%%%%%%%%%%%%%%%%%%%%%%%%%%%%%%%

The theory greatly simplifies when $\alpha =90{{}^\circ}$ (the
vertical pannel case). Indeed, in that case,
\beq
X\left( \theta _{0},\pi \right)
=-X\left( \theta _{0},0\right) =-r_{11}^{\prime }\left( \theta _{0}\right) 
\eeq
where $r_{11}^{\prime }\left( \theta _{0}\right) $ is given by (\ref{r'11}).
Solving $B_{1}\left( \theta _{0}^{*}\right) =B_{2}\left( \theta
_{0}^{*}\right) $ in that case, yields $\tan \left( 2\theta _{0}^{*}\pm
\left( \theta _{1}+\theta _{2}\right) \right) =0$, which
yields $\theta _{0}^{*}=\left( \theta _{1}+\theta _{2}\right) /2$,
corresponding to the choice `-;-' because $\cos \left( \theta _{1}-\theta
_{0}^{*}\right) >0$ and $\cos \left( \theta _{2}-\theta _{0}^{*}\right) >0$,
meaning $\theta _{2}-\theta _{1}^{{}}<\pi.$
The optimal radius of the patch is
then
\[
r_{*}=R_{2}\left( \theta _{0}^{*}\right) \sin \theta _{0}^{*}=R_{1}\left( \theta _{0}^{*}\right) \sin \theta _{0}^{*}
\]
where the expression of $R_{2}\left( \theta _{0}^{*}\right) $ is found in
(\ref{eq:bond})(\ref{tan2}). In Fig. \ref{cond}, we plot the optimal radius $r_{*}$ for
various values of $\left( \theta _{1},\theta _{2}\right) $.

Then, up to the constant $K(S)$ in (\ref{K}), the
optimal volume thus stored is $V_{2}(r)/r^{2}$ given by (\ref{V:r2}),
 evaluated at $\theta _{0}^{*}$. In Fig. \ref{Vr2}, we plot the optimal volume stored  for
various values of $\left( \theta _{1},\theta _{2}\right) .$

In some extreme cases, the maximum of $V_{2}(r)/r^{2}$ can be
reached at some $r<r_{*}.$
Taking indeed the derivative of $V_{2}(r)/r^{2}$ with
respect to $x_{0}:=\cos \theta _{0}$ and forcing this derivative to vanish
when $x_{0}^{*}:=\cos \theta _{0}^{*}$ (with $\theta _{0}^{*}=\left( \theta
_{1}+\theta _{2}\right) /2$) yields a domain in the $\left( \theta _{1},\theta
_{2}\right) -$plane separating a region where the optimal radius should be
computed at the tip of the medallion (when overflow and dewetting occur
simultaneously), from a small region where the maximum is found on the overflow
curve. The equation of the separating line is given by the following: 
\beq
B_{2}\left( \theta _{0}\right) =\frac{\tan \left( \theta _{0}-\theta
_{2}\right) }{r_{11}^{\prime }\left( \theta _{0}\right) }\approx \frac{\tan
\left( \theta _{0}-\theta _{2}\right) }{x_{0}-2/\left( 1+x_{0}\right) }
\eeq
\beq
=\frac{\tan \left( \theta _{2}-\theta _{0}\right) \left( 1+x_{0}\right) }{%
\left( 1-x_{0}\right) \left( 2+x_{0}\right) },
\eeq
we have 
\beq
\frac{V_{2}\left( r\right) }{r^{2}} \approx \left( B_{2}\left( \theta
_{0}\right) \right) ^{1/2}\frac{\left( 1-x_{0}\right) ^{2}\left(
2+x_{0}\right) }{1-x_{0}^{2}}
\eeq
\beq
=\left( B_{2}\left( \theta _{0}\right) \right)
^{1/2}\frac{\left( 1-x_{0}\right) \left( 2+x_{0}\right) }{1+x_{0}} 
\eeq
\beq
=\left[ \frac{\tan \left( \theta _{2}-\theta _{0}\right) \left(
1-x_{0}\right) \left( 2+x_{0}\right) }{1+x_{0}}\right] ^{1/2}=A\left(
x_{0}\right) ^{1/2}
\eeq
where
\beq
A\left(x_{0}\right)=\frac{\left( 2x_{0}\left( 1-x_{0}^{2}\right) ^{1/2}-\sin \left( 2\theta
_{2}\right) \right) \left( 1-x_{0}\right) \left( 2+x_{0}\right) }{\left(
x_{0}^{2}-\sin ^{2}\theta _{2}\right) \left( 1+x_{0}\right) }
\eeq
Solving
\beq
\frac{dA}{dx_{0}}\Big|_{\theta _{0}^{*}=\left( \theta _{1}+\theta
_{2}\right) /2}=0 
\label{line}
\eeq
yields two separating lines shown in Fig. \ref{cont}. In the extreme shaded regions (superhydrophilic or superhydrophobic, unrealistic), the optimal radius of the patch is found on the overflow curve and not at the tip of the medallion. Elsewhere and so in practice, in the white region, the optimal radius of the patch is found when simultaneous dewetting and overflow apply.

%%%%%%%%%%%%%%%%%%%%%%%%%%%%%%%%%
\begin{figure}[tbp]
\begin{center}
\includegraphics[width=0.45\textwidth]{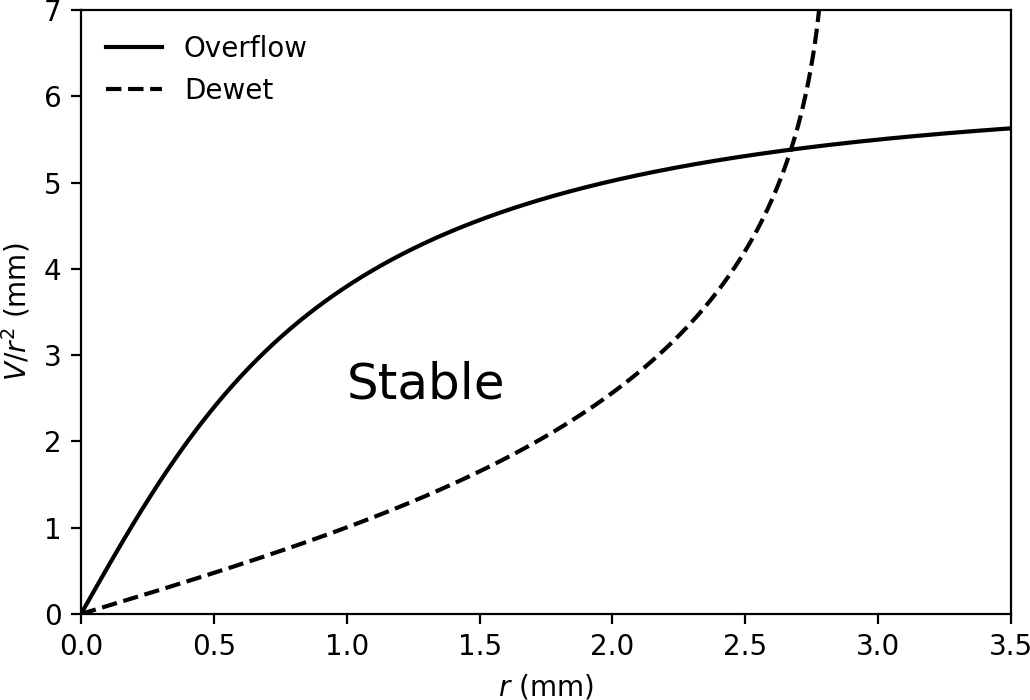} 

\end{center}
\caption{$V/r^2$ function of $r$ at overflow (continuous line)
and at dewetting (dashed line) for $\theta_1=57^\circ$ and $\theta_2=120^\circ$. Slope $\alpha=90^\circ$.}
\label{stable}
\end{figure}
%%%%%%%%%%%%%%%%%%%%%%%%%%%%%%%%%

In Fig. \ref{stable}, we plot $V/r^2$ as a function of $r$ at dewetting (continuous line)
and at overflow (dashed line) for $\theta_1=57^\circ$ and $\theta_2=120^\circ$, for a slope $\alpha=90^\circ$.

Optimal storage is at the intersection ($r_*,V_*/r_*^2$) of the two curves. The region on the right of the dashed curve is unstable by dewetting. The region above the continuous curve is unstable by overflow. Stability holds only inside the medallion. If $r>r_*$, dewetting is unavoidable.
If $r$ is a little below $r_*$, storage is a little below the optimal value along the continuous curve (overflow).

\subsection{Results varying $\al$}

\begin{figure}[tbp]
\begin{center}
\includegraphics[width=0.45\textwidth]{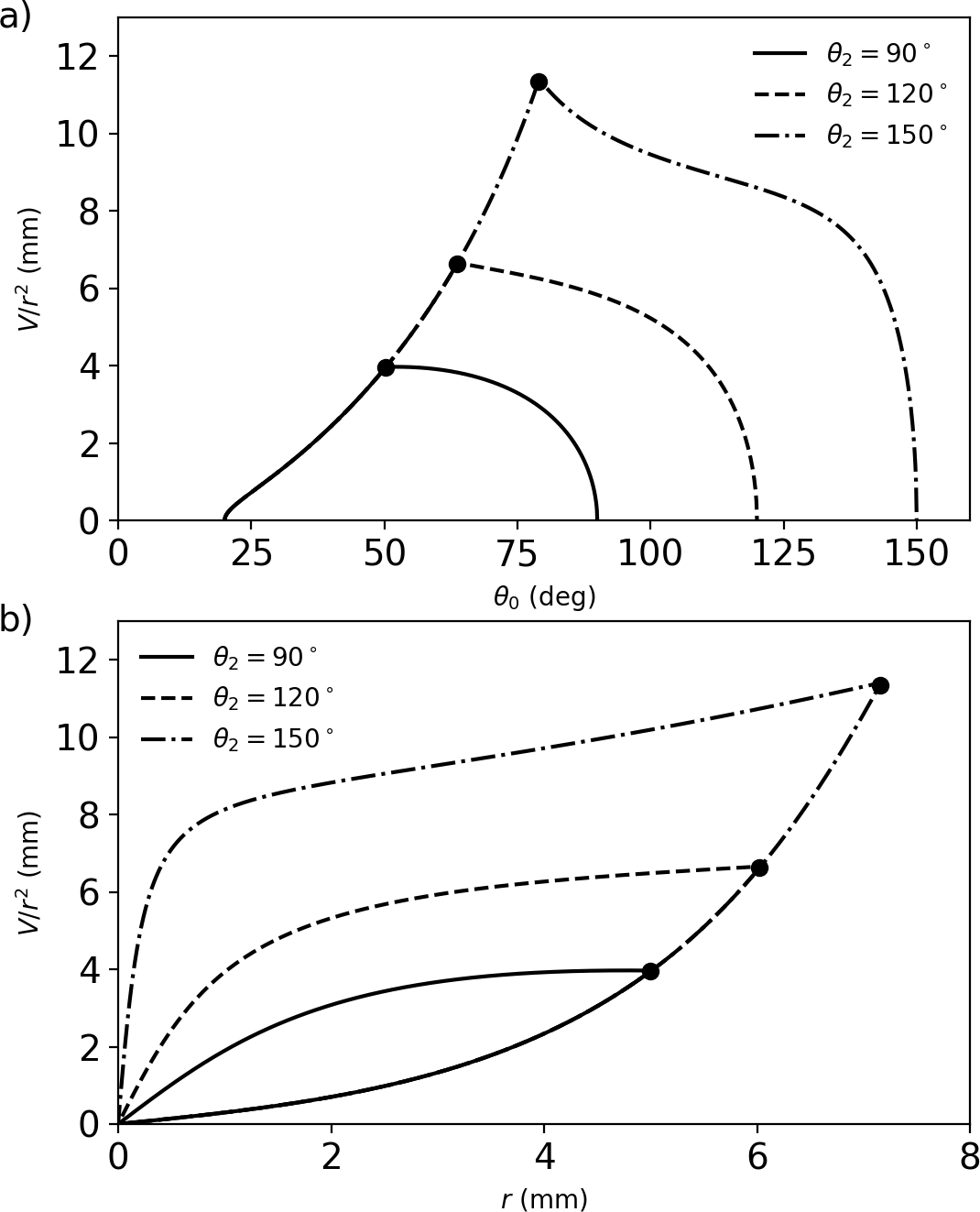} 
\end{center}
\caption{a) $V(r)/r^2$ versus $\theta_0$ for different values of $\theta_2$
and $\alpha = 40^\circ$ and $\theta_1=30^\circ$. b) $V(r)/r^2$ versus $\mathrm{r}$ for different
values of $\theta_2$ and $\alpha = 40^\circ$ and $\theta_1=20^\circ$.}
\label{fig3}
\end{figure}

In Fig. \ref{fig3}a) we give a plot of $V(r)/r^{2}$ against $\theta_{0}$,
calculated from (\ref{V:r2}) for $\alpha =40^{\circ }$, $\theta_1=30^\circ$, for
different values of $\theta_{2}$. We used $\gamma =0.072$ J/m$^{2}$ and $%
\rho =997$ kg/m$^{3}$ (standard values for water). In all cases, at $\theta
_{0}=\theta_{0}^{*}$ the curves present a maximum indicating the maximum
storage capacity of the surface before the roll-off of a single drop. Fig. 
\ref{fig3}b) represents a parametric plot of $V(r)/r^{2}$\ against $r$,
using $r=R\left( \theta_{0}\right) \sin \theta_{0}.$ The optimal radius $%
r_{*}$ of the patch maximizing $V(r)/r^{2}$ can directly be read.

%%%%%%%%%%%%%%%%%%%%%%%%%%%%%%%%%
%%%%%%%%%%%%%%%%%%%%%%%%%%%%%%%%%

\begin{figure}[tbp]
\begin{center}
\includegraphics[width=0.45\textwidth]{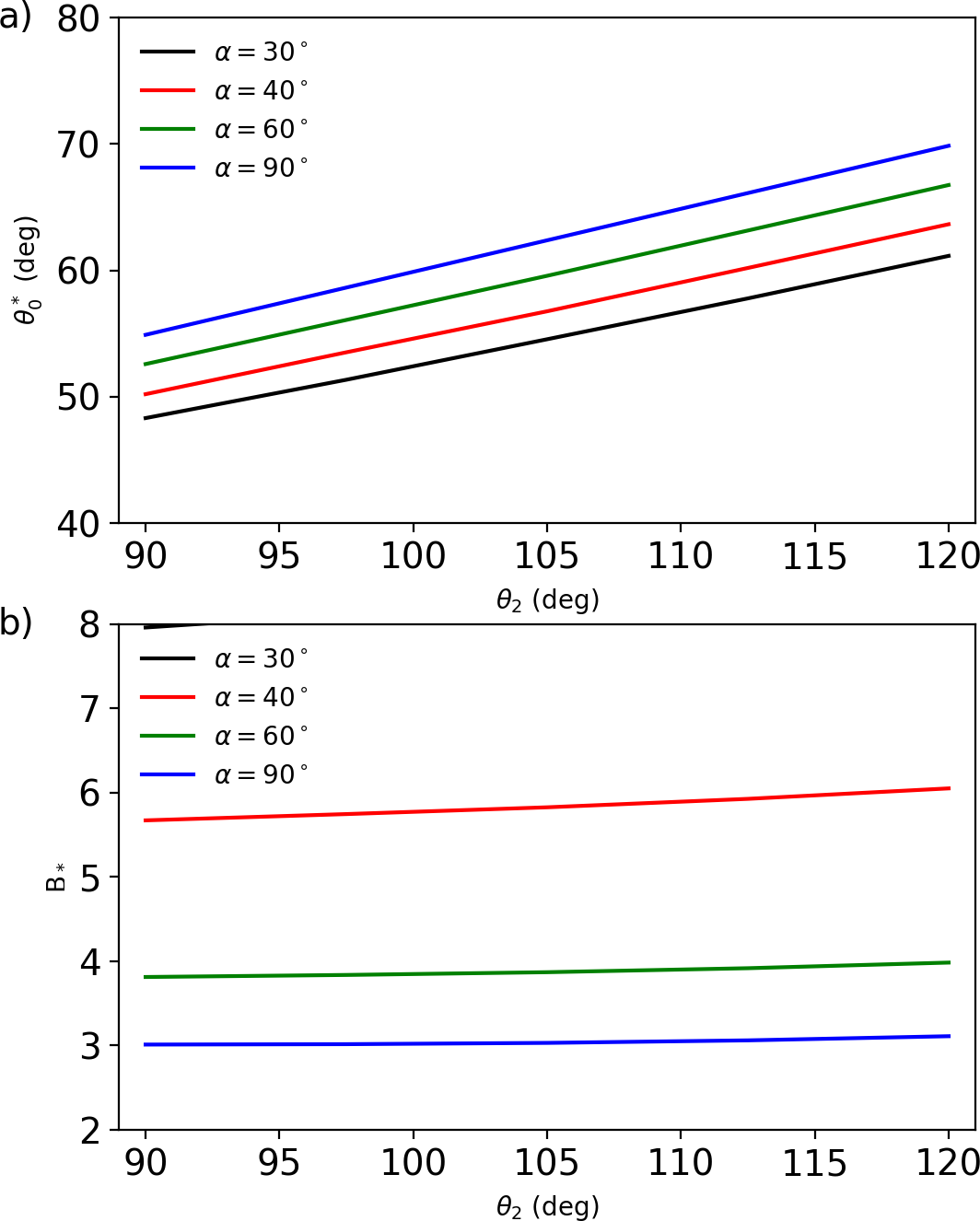} 
\end{center}
\caption{a) $\theta_{0}^{*}$ and b) Bond number associated to the maximum $%
V/r^2$ versus $\theta_2$ for the different values of $\alpha$. In both cases $\theta_1=20^\circ$.}
\label{fig4}
\end{figure}

Fig. \ref{fig2} shows the drop profile which maximizes $V(r)/r^{2}$ for $%
\theta_{2}=150^{\circ }$, $\alpha =40^{\circ }$ and $\varphi =0, \pi$ as
computed from (\ref{ansatz})(\ref{V:r2}). We can also determine the location of $%
\theta_{0}^{*}$ for $\theta_{2}\in [20^{\circ },180^{\circ }]$ and for
different values of $\alpha $. This is done in Fig. \ref{fig4}a) considering $\theta_1=20^\circ$.\emph{\ }

Once we obtained $\theta_{0}^{*}$, we calculate the optimal Bond number $%
B_{*}=B\left( R\left( \theta_{0}^{*}\right) \right) $. This is shown in
Fig. \ref{fig4}b), using (\ref{bond}). As has been shown in \cite
{DDH17}, the small Bond number approximation for the Laplace-Young's
equation remains within 1\% of deviation with respect to the exact solution
computed numerically with \textit{surface evolver} until $B$ of order 10 and
for $\alpha =30^{\circ }$. Therefore, we expect this approximation to be
reasonably close to the exact solution for $B<10$ which is verified for the range of $\theta_1$, $\theta_2$ and $\alpha$ considered in the present work. 

%%%%%%%%%%%%%%%%%%%%%%%%%%%%%%%%%
%%%%%%%%%%%%%%%%%%%%%%%%%%%%%%%%%

\begin{figure}[tbp]
\begin{center}
\includegraphics[width=0.45\textwidth]{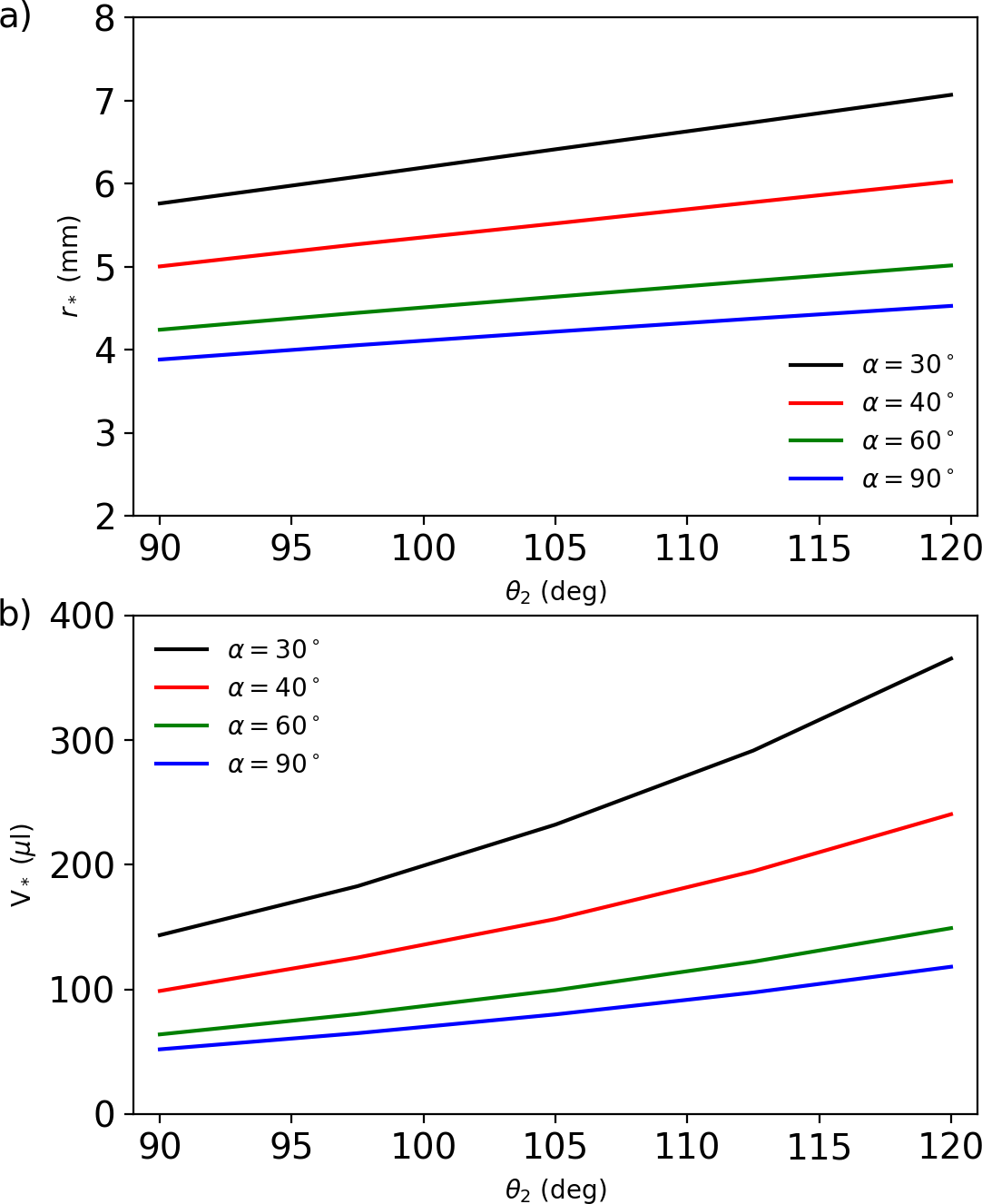} 
\end{center}
\caption{a) Optimal radius of the patch and b) Single patch volume,
associated to the maximum $V/r^{2}$ against $\theta_{2}$, for different
values of $\alpha $, for $\theta_1=20^\circ$.}
\label{fig5}
\end{figure}

The optimal radius $r_{*}$ of the patch which maximizes the storage capacity
of the surface is calculated with $r_{*}=R(\theta_{0}^{*})\sin \theta
_{0}^{*}$ where $R(\theta_{0}^{*})$ is computed from (\ref{eq:bond}). The
resulting values are shown in Fig. \ref{fig5}a) for various values of $%
\alpha $ and $\theta_1=20^\circ$. For $\alpha \in [30^{\circ },90^{\circ }]$ and values of the
contact angle in the hydrophobic region ${90^{\circ }<\theta_{2}<120^{\circ }}$, we see
that the optimal values $r_{*}$ for the hydrophilic patch are in the range $%
[3-7]$ mm. 

Fig. \ref{fig5}b) shows the optimum volume per patch $V_{*}:=V\left(
r_{*}\right) $ maximizing the global water storage versus $\theta_{2}$,
for different $\alpha $.

\begin{figure}[tbp]
\begin{center}
\includegraphics[width=0.45\textwidth]{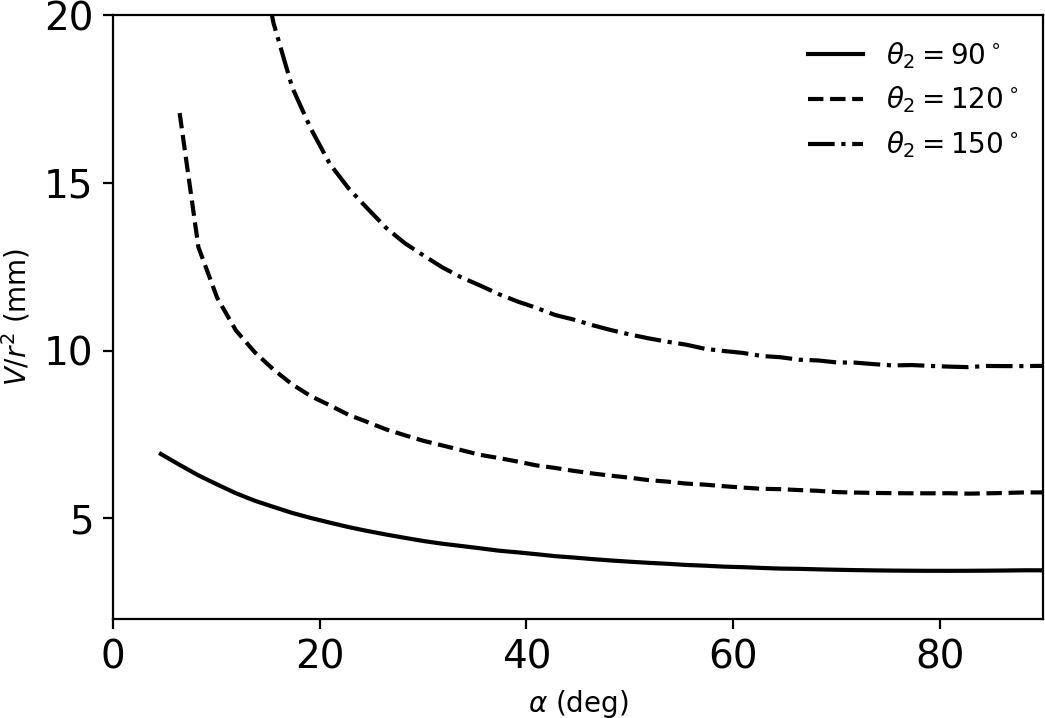} 
\end{center}
\caption{$V(r_{*})/r_{*}^{2}$ against $\alpha $, for different values of $%
\theta_{2}$, for $\theta_1=20^\circ$.}
\label{fig6}
\end{figure}

Finally, in Fig. \ref{fig6}, we plot $V(r_{*})/r_{*}^{2}$ against $\alpha $
for various values of $\theta_{2}$ and for $\theta_1=20^\circ$. As expected, it is a decreasing
function of $\alpha\in(0,\pi/2)$. %%%%%%%%%%%%%%%%%%%%%%%%%%%%%%%%%

%%%%%%%%%%%%%%%%%%%%%%%%%%%%%%%%%

%%%%%%%%%%%%%%%%%%%%%%%%%%%%%%%%%

%\newline

\begin{figure}[tbp]
\begin{center}
\includegraphics[width=0.48\textwidth]{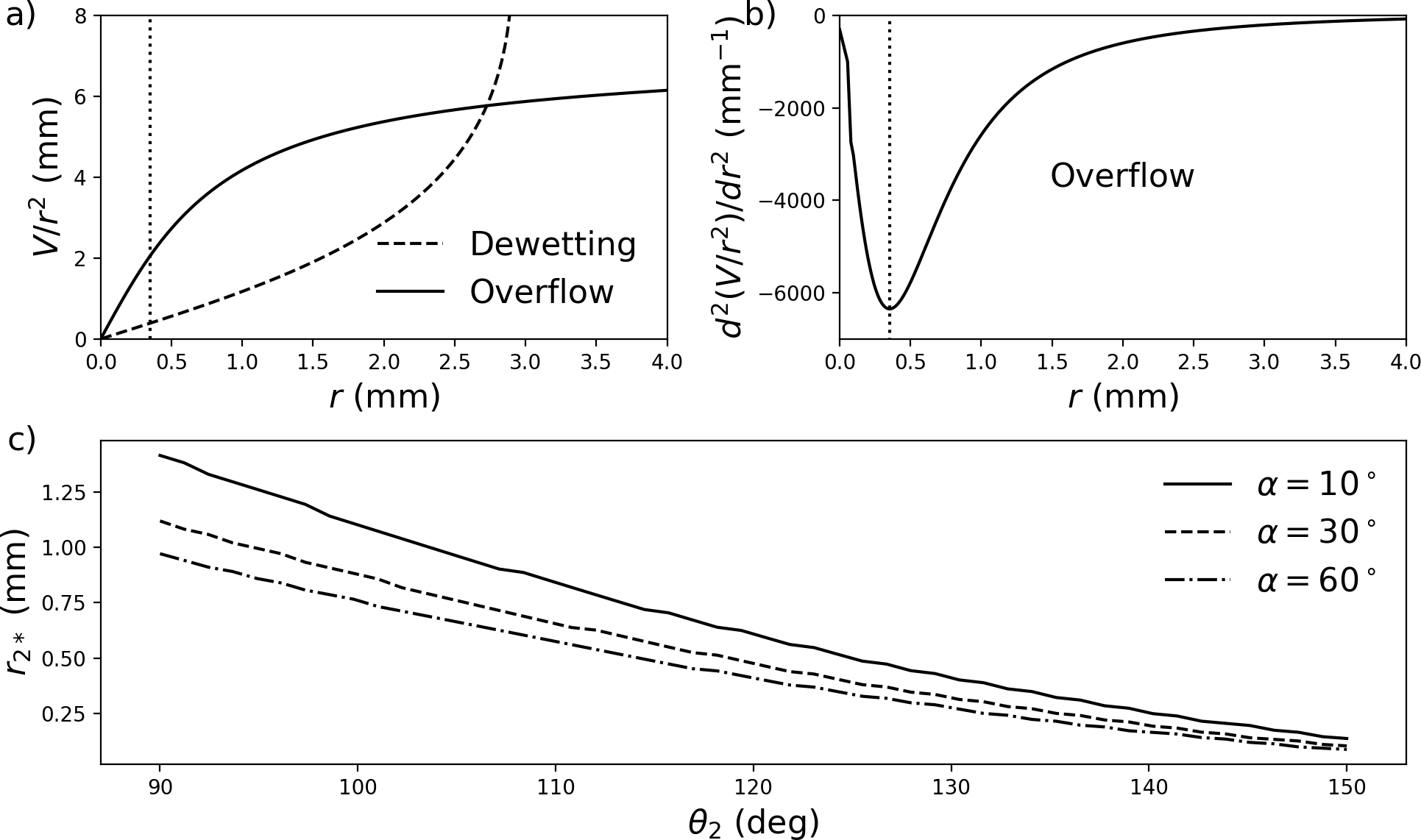} 
\end{center}
\caption{a) $V/r^2$ curves for $\theta_1=64^\circ$, $\theta_2=124^\circ$ and $\alpha=60^\circ$. b) $d^2(V/r^2)/dr^2$ for overflow, for the same parameters. The minimum is located at $r_{2*}$ c) $r_{2*}$ versus $\theta_2$ for different tilting angles.}
\label{figrSteno}
\end{figure}

Let us end with two illustrative examples:

- The stenocara beetle example, roughly corresponding to $%
\alpha =60{^{\circ }}$,\, $\theta_1=64^\circ$ and $\theta_{2}=124{^{\circ }}$, see \cite{Guadarrama14}.
A tilt angle is necessary for
the fog drops to strike the back of the beetle and also to collect dew
or fog water by gravity.
We assume the
beetle's back area to be about $S=200$\ mm$^{2}$. From our calculations, $\delta
=r_{*}=2.7$ mm which is a little larger than the expected one. However, the corresponding curve of $V_2/r^2$ versus $r$ in the overflow case, shown in Fig. \ref{figrSteno}a), presents a fast increase of the volume per surface area for small $r$ but this variation becomes slower as $r$ increases. In fact, the second derivative of $V_2/r^2$ for overflow, shown in Fig. \ref{figrSteno}b), presents a minimum at $r_{2*}$ that characterizes this transition.  In Fig. \ref{figrSteno}c) we represent the location of $r_{2*}$ versus $\theta_2$ for different tilting angles where we observe that $r_{2*}<0.5$ mm for $\theta_2>110^\circ$. In particular, for $\alpha=60^\circ$ and $\theta_2=124^\circ$ we have $r_{2*}=0.35$ mm corresponding to $V\left( r_{2*}\right) =0.25$\ mm$^{3}$ and, from (\ref{Numb}),
\begin{eqnarray*}
N\left( r_{2*}\right) &=&\frac{2S}{\sqrt{3}\left( 3r_{2*}\right) ^{2}}\simeq
208 \\
N\left( r_{*}\right) V\left( r_{*}\right) &=&K\left( S\right) V\left(
r_{*}\right) /r_{*}^{2}\simeq 53 \text{\,\rm mm}^{3}
\end{eqnarray*}

%%%%%%%%%%%%%%%%%%%%%%%%%%%%%%%%%
%%%%%%%%%%%%%%%%%%%%%%%%%%%%%%%%%

The beetle's back is made of about five hundred hydrophilic patches and a pannel
of such patches with surface $S=200$\ mm$^{2}$ would collect about $75$ mm$%
^{3} $ if it is surrounded by a strongly hydrophobic matrix. The beetle could thus carry $75$\ mg on its back,
about $1/30$ of its weight.\emph{\newline
}

- Let us consider a water-collecting problem at a different spatial scale.
Consider a pannel of surface $S=200$\emph{\ }cm$^{2}$ with inclination $%
\alpha =30{^{\circ }}$\emph{\ }, $\theta_1=30^\circ$, and\emph{\ }$\theta_{2}=106{^{\circ }}$
(much less hydrophobic than in the beetle's case)\emph{. }From our
calculations, $r_{*}=3.7$\ mm$,$\ $\delta =r_{*},$\ $V\left( r_{*}\right)
=72$\ mm$^{3}$, leading to 
\begin{eqnarray*}
N\left( r_{*}\right) &=&\frac{2S}{\sqrt{3}\left( 3r_{*}\right) ^{2}}\simeq
170 \\
N\left( r_{*}\right) V\left( r_{*}\right) &\simeq&14\text{\,\rm cm}^{3}
\end{eqnarray*}
This shows that a hydrophobic pannel of $S=200$\ cm%
$^{2}$ having a contact angle $\theta
_{2}=106{^{\circ }}$ with hydrophilic patches of $\theta_1=30^\circ$  would approximately collect $14$ cm$^{3}.$
We can refer to the experimental results described in  \cite{Garrod07} where the use of purely homogeneous surfaces is compared to heterogeneous surfaces.  The results indicate that surfaces with patches can collect significantly more water than homogeneous surfaces. 

\section{Dynamics of water harvesting}
It is known that patterning will promote dropwise condensation if associated with a thermal cooling of the substrate. The above calculations prove that there is an optimal size of the patches to store water under static conditions, for any wettabilities. Now, the key question is: can one relate these static results to fog harvesting or water adsorption occurring under dynamical conditions.
If we introduce time in the problem, to study for instance condensation or fog harvesting, we thus have to consider different time scales: 

- $t_1$, the time required for the adsorbed liquid molecules to cover the hydrophilic area of the plate, assumed approximately independent of $r$.

- $t_2$, the time to fill the patch up to the maximum $V(r)$. 

- $t_3$, the time for removal of the liquid from the surface, typically within cascades. The time $t_3$ is expected to be small compared to $t_1$ and $t_2$.

We are dealing with a mono-disperse array of patches, expected to synchronize.   
The time to fill a hydrophilic patch up to the maximum filling $V(r)$ can thus be written as 
\begin{equation} \label{time}
 t_{\mathrm{fill}}\approx t_1+{\frac{V(r)}{\pi r^2(av+c)}} 
\end{equation}
where the second term of the right-hand side is $t_2$. In this expression, $v$ is the fog velocity, $a$ is a volume fraction of fog adsorption onto the substrate, and $c$ accounts for condensation occurring
also when $v=0$. If, for a simple example, $v=$ 1 m/s, $a$=0.01, $c=0$, $t_1=t_2$ and $r=$1 mm,  (\ref{time}) yields $t_{\mathrm{fill}}$ of order 0.1 seconds.

Note however that (\ref{time}) is designed to give the $r$ dependence of $t_{\mathrm{fill}}$. There remains a global time factor not predicted by the model.

The volume of water collected per unit time and unit area, neglecting correlations, is
then 

%\[
%Q(r)={\frac{V(r)}{t_{\mathrm{fill}}}}{\frac{$K\left( S\right){r^2}}\approx {\frac{%
%V(r)/r^2}{t_1+b\,V(r)/r^2}} 
%\]
$$
Q(r)={K(S)V(r)/r^2\over t_1+V(r)/(\pi r^2(av+c))}
$$
It appears that $Q(r)$ is maximum when $V(r)/r^2$ is
maximum: statics and dynamics agree. 

\section{Experimental results}

To compare with experiment, we need to prepare patches of different radii $%
r$ and to deposit several water droplets of a certain volume (usually $10$
microliters but it can be $5$ or $2$ microliters) and for each deposition we
tilt the surface to reach $\alpha $. Herewith we will concentrate on $%
\alpha =90{{}^\circ}$ and $45{{}^\circ}$ but the results can easily be
extended to other values. For this experiment we proceed as follows: we deposit different
drop volumes of water inside the patch using a syringe and we tilt the
surface to the corresponding inclination $\alpha $ to observe if the drop
remains on the patch or if it leaves the patch. For each drop volume, we
measure the contact radius $r_c$ and the two lateral contact angles $\theta^{\min }$ and $\theta^{\max }$ versus the
volume for a tilting angle of $\alpha =0$, $45$ and $90$ degrees.

Different techniques may be considered for the patterning of a surface.
Herewith we know from biomimicry that due to gravity our best patterns
should have a radius of the mm size, the technique may thus be based on
classical methods such as micro-lithography, laser patterning, 3D printing,
contact printing, roll coating on top of a mask. Our result is not limited
by the considered technology.

For the sake of simplicity, we realize here these surfaces by first grafting
hydrophobic OTS molecules on a $3$ cm x $3$ cm cleaned glass substrate using a
standard procedure, \cite{Bourdon13}. Once the coatings are realized and validated (by
contact angle measurements), we then realize the circular patches through a
PTFE mask with holes of a given radius and, using UV-O3 radiation (UVOCs oven
from Applitek), we produce hydrophilic glassy patches. The contact angles of the hydrophilic patches (subscript $p$) and the hydrophobic matrix (subscript $m$) are shown in Table \ref{tab_ang}.

\begin{table} 
%\label{tab_ang}
\begin{tabular}{ccc} 
\\ 
\hline
Hydrophobic matrix &  &  \\
$\theta _{m}^{0}$ (deg)& $\theta _{m}^{adv}$ (deg)& $\theta _{m}^{rec}$ (deg)\\
$115.3\pm 3.8$ & $118.9\pm 4.7$ & $105.1\pm 3.7$\\
\\ \hline
Hydrophilic patches&&\\
$\theta_{p}^{0}$ (deg)& $\theta _{p}^{adv}$ (deg)& $\theta _{p}^{rec}$ (deg) \\
 $26.3\pm 4.83$ & $%
48.3\pm 1.0$ & $19.9\pm 3.3$
\\ \hline
\end{tabular}
%\label{tab_ang}
\caption{Experimental measurements for the equilibrium, advancing and receding contact angles of both the hydrophilic patches and the hydrophobic matrix.}
\label{tab_ang}
\end{table}

\begin{figure}[tbp]
\begin{center}
\includegraphics[width=0.45\textwidth]{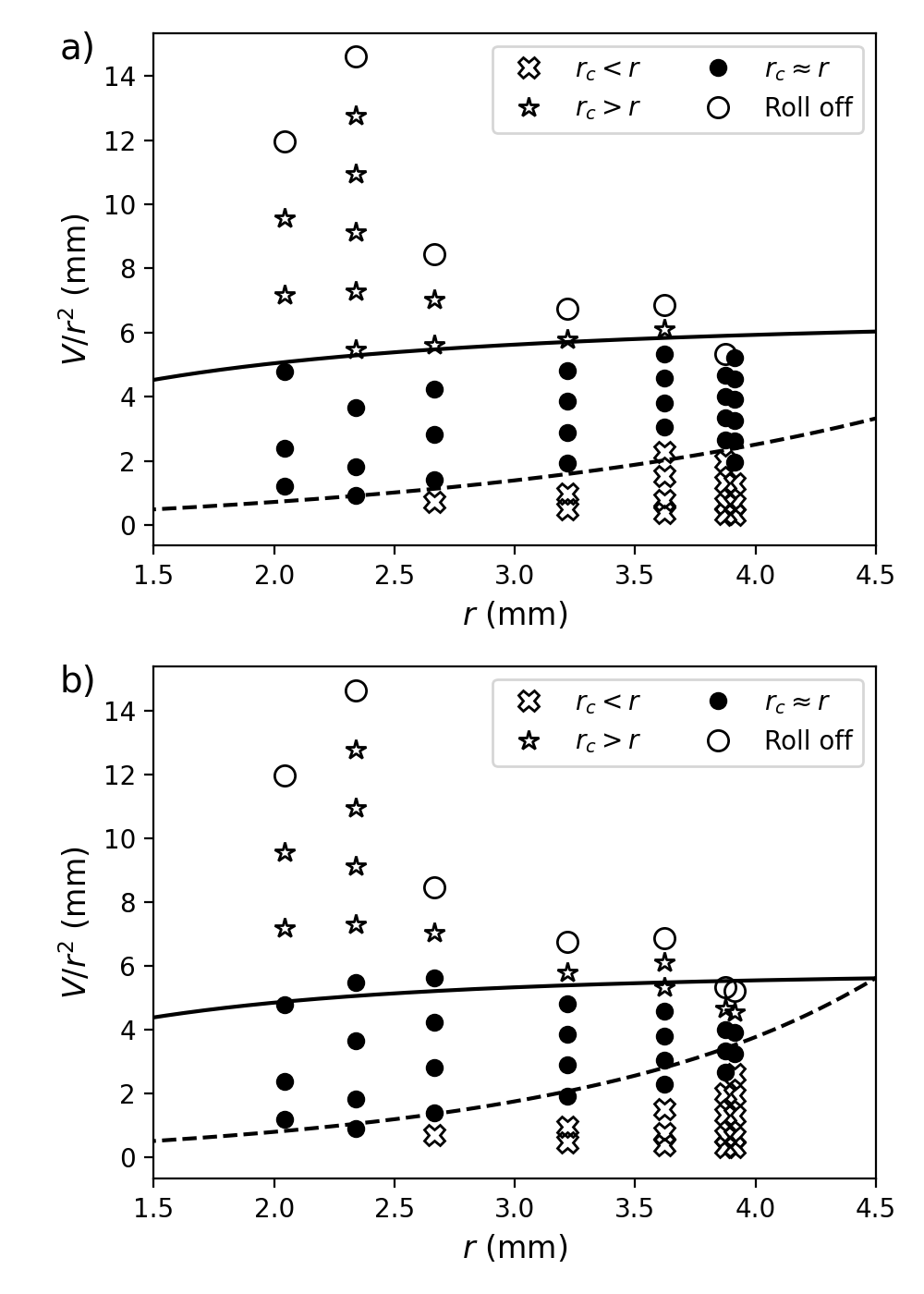} 
\end{center}
\caption {The stored volume of water per unit area on the surface versus patch radius for a) $\alpha
=45{{}^\circ}$ and b) $\alpha =90{{}^\circ}$ for the glass/OTS system. The continuous (dashed) line is the theoretical one of ${V_{2}(r)}/{r^{2}}$ (respectively  ${V_{1}(r)}/{r^{2}}$) versus $r$ given in Section II.A and plotted in Fig. \ref{fig3}b) for different sets of parameters.}
\label{figExp}
\end{figure}

After the UV treatment, we characterize the surfaces by water contact angle
measurements.
The water drop increment volume for the characterization is $5$ microliters.
The advancing and receding contact angles are measured with the Johnson and
Dettre method. Figure \ref{figExp} shows the ratio $V/r^2$ versus $r$ where each point corresponds to a deposited droplet volume $V$ at the center of the patch  for tilting angles 
$\alpha =45{{}^\circ}$ a) and $90{{}^\circ}$ b). The points can be classified in four different groups:

\begin{itemize}
    \item $r_c<r$ (contact radius smaller than the patch radius). The drop behaves as on a homogeneous substrate. The contact radius increases with the volume meanwhile the contact angle remains approximately constant.
    
    \item $r_c\approx r$ (contact radius close to the patch radius). The contact line of the drop is pinned on the edge of the patch. In this region, the contact radius remains constant and the contact angle depends on the drop volume. 
    
    \item $r_c>r$ (contact line depins partly).  The contact line overpasses the patch boundary but the drop remains pinned.
    
    \item Roll-off. The drop rolls down from the plate.
\end{itemize}

The boundary lines between the different regions (as sketched in Fig.~\ref{figExp}) are the theoretical ones of ${V_{2}(r)}/{r^{2}}$ and  ${V_{1}(r)}/{r^{2}}$ versus $r$, as given in Section II.A; see also Fig. \ref{fig3}b) of Section II.C.
The continuous and dashed lines of Fig. \ref{figExp} correspond respectively to the theoretical overflow and dewet curves predicted by the model for  $\theta _1 =  \theta _{p}^{rec} $ and $\theta _2 =  \theta _{m}^{adv} $. All experimental points on the medallion between the theoretical curves corresponds to drops with the contact line pinned at the edge of the patch ($r_c\approx r$) in agreement with the theoretical model. The points outside the medallion cannot correspond to an optimal situation. Drops under the dewetting curve cannot fill the whole patch and depin from the top of the patch.   Drops above the overflow curve will overpass the patch, covering part of the hydrophobic matrix.  The stability medallion appearing in Fig. \ref{figExp} thus qualitatively confirms the theoretical predictions of Figs.~\ref{stable} and~\ref{figrSteno}(a).

\section{Summary}

Inspired by the water-harvesting
strategy of the stenocara beetle in the Namibian desert, we considered a water collection device comprising a patterned surface inclined by an angle $\alpha$ with the horizontal. The patterned surface is made of hydrophilic patches characterized by a contact angle $\theta_1$ embedded in a hydrophobic surface characterized by a contact angle $\theta_2$. The hydrophilic patches are approximately circular with a radius $r$ and arranged according to a triangular pattern, approaching close-packing. We found that, except for explicit unrealistic values of ($\theta_1$, $\theta_2$) (see Fig. \ref{cont}), for each value of the tilt angle $\alpha$, there is an optimal radius of the patches maximizing the volume stored per unit surface area and that it is found when simultaneous dewetting and overflow occur at the back and the front of the drop respectively. Experiments with a glass/OTS system confirmed this prediction and the existence of a stability medallion. 
The analytical results were obtained using the small Bond number analysis of the non-axisymmetric Young-Laplace equation developed in \cite{DDH17}. 
It is argued, with a simple dynamical model, that the volume of water collected per unit time and unit area is maximum when $V(r)/r^2$ is maximum, like in the static case.
%The distance r is such that for an increasing water droplet pinned to a hydrophilic %patch with $\theta^{\max }$ being the contact angle at the downhill front of the %droplet and $\theta^{\min }$ being the contact angle at the uphill back of the %droplet, and for the relevant angle $\alpha$, $\theta^{\max }$ exceeds $\theta_2$ and %$\theta^{\min }$ passes below $\theta_1$ simultaneously.
Our results are valid for any combination of the angles $\theta_1$ and $\theta_2$ provided that $\theta_1 < \theta_2$, and can thus be applied to two hydrophobic surfaces, one less than the other, avoiding hydrophilic surfaces known to be difficult to maintain over time. The results were applied on one hand to the stenocara beetle and on the other hand to a model of a water-collecting system at a larger spatial scale, leading to data with realistic magnitude.

%Considering the problem of describing the shape and volume of  a 
%sessile drop sitting on an incline with angle $\alpha $ and with circular
%footprint, we obtained the exact solution in spherical %coordinates of the
%Laplace-Young equation at first order in the Bond number, a reasonably fair
%Inspired by the water-harvesting
%strategy of the stenocara beetle in the Namibian desert, we considered an
%ideal regular array of hydrophilic (contact angle $\theta_{1}$) circular
%patches of radius $r$ embeddded in a hydrophobic (contact angle $\theta_{2}$%
%) medium, sat on an incline of tilt $\alpha $ with respect the horizontal.
%We have shown that the total liquid storage capacity of this structure on a
%given surface (as represented by $V(r)/r^{2}$) can be maximized by selecting
%an optimal radius of the patches whose dependence on $\theta_{2}$, $\theta_{1}$ and %$\alpha $ was elucidated. 

\bigskip

{\bf Acknowledgments:}
Special thanks to A. Draux for stimulating discussions. The authors also thank the European Space Agency (ESA), France
and the Belgian Federal Science Policy Office (BELSPO) for their
support in the framework of the PRODEX Programme. This research
was  partially funded by  FNRS and R\'egion Wallonne. 
%\end{acknowledgments}

%\clearpage
\bibliographystyle{apsrev4-1}
\bibliography{ddh17}

\end{document}